\documentclass[prb,onecolumn,12pt]{revtex4}
\usepackage{epsfig}
\usepackage{amssymb}
\usepackage{threeparttable}
\usepackage{graphicx}
\usepackage{longtable}

\begin{document}
\title{Thermal conductivity of monolayer MoS$_2$, MoSe$_2$, and WS$_2$: Interplay of mass effect, interatomic bonding and anharmonicity}
\author{Bo Peng$^1$, Hao Zhang$^{1,*}$, Hezhu Shao$^2$, Yuchen Xu$^1$, Xiangchao Zhang$^1$ and Heyuan Zhu$^1$}
\affiliation{$^1$Shanghai Ultra-precision Optical Manufacturing Engineering Center, Department of Optical Science and Engineering, Fudan University, Shanghai 200433, China\\
$^2$Ningbo Institute of Materials Technology and Engineering, Chinese Academy of Sciences, Ningbo 315201, China}

\begin{abstract}
Phonons are essential for understanding the thermal properties in monolayer transition metal dichalcogenides, which limit their thermal performance for potential applications. We investigate the lattice dynamics and thermodynamic properties of MoS$_2$, MoSe$_2$, and WS$_2$ by first principles calculations. The obtained phonon frequencies and thermal conductivities agree well with the measurements. Our results show that the thermal conductivity of MoS$_2$ is highest among the three materials due to its much lower average atomic mass. We also discuss the competition between mass effect, interatomic bonding and anharmonic vibrations in determining the thermal conductivity of WS$_2$. Strong covalent W-S bonding and low anharmonicity in WS$_2$ are found to be crucial in understanding its much higher thermal conductivity compared to MoSe$_2$.
\end{abstract}

\maketitle

\section{Introduction}

Monolayer transition metal dichalcogenides, MX$_2$ (M = Mo, W; X = S, Se), have aroused much interest recently due to their remarkable properties for applications in next-generation nanoelectronic and energy conversion devices \cite{C4NR01600A,Klinovaja2013,Photothermoelectric-MoS2}. Since all these applications are closely related to its thermal properties, it is necessary to investigate the lattice dynamics and thermodynamic properties of MX$_2$. For instance, high-performance electronic devices strongly depend on high thermal conductivity for highly efficient heat dissipation, while low thermal conductivity is preferred in thermoelectric application.

In semiconductors, heat is carried by the atomic vibrations that are quantized as phonons \cite{Ziman}. Theoretical predictions based on the phonon Boltzmann transport equation have found that monolayer WS$_2$ has the highest thermal conductivity (142 W/mK) at room temperature, then followed by MoS$_2$ (103 W/mK) and MoSe$_2$ (54 W/mK) \cite{Gu2014}. However, the measured thermal conductivity for monolayer MoS$_2$ and WS$_2$ is $34.5\pm4$ W/mK \cite{monoMoS2-k} and 32 W/mK \cite{raey}, respectively, which is much lower than the theoretical predictions. Furthermore, various phonon properties such as interatomic bonding and anharmonic vibrations still lack a unified understanding. The parameters that affect phonon transport include crystal structure, atomic mass, interatomic bonding, and anharmonicity \cite{Slack1973321,Lindsay2013,Jain2014}. Generally there are four rules for finding a nonmetallic crystal with higher thermal conductivity, including (i) lower average atomic mass, (ii) stronger interatomic bonding, (iii) simpler crystal structure, and (iv) lower anharmonicity \cite{Slack1973321}. All monolayer MX$_2$ compounds have similar crystal structures, while conditions (i) and (ii) imply a larger Debye temperature, and condition (iv) means smaller Gr\"uneisen parameter. Recent theoretical investigation has provided a quantitative analysis of the roles of mass, structure, and bond strength in thermal expansion and thermomechanics of MX$_2$ \cite{Huang2015}. However, the roles of mass, interatomic bonding, and anharmonic vibrations in phonon transport still remain uninvestigated. Clear knowledge of the underlying physics will be helpful for understanding and modulating the thermal transport in MX$_2$, for example, through doping other M or X atoms \cite{Gong2014,Dumcenco2013}.

Here we investigate fundamental vibrational properties to understand thermal transport in MoS$_2$, MoSe$_2$, and WS$_2$. The measured phonon frequencies are well reproduced in our calculations. The thermodynamic properties are calculated within quasi-harmonic approximation, and the calculated thermal conductivities agree well with the measurements. Combining first principles calculations and the Slack model, the roles of mass, interatomic bonding, and anharmonicity in thermal transport are clearly revealed.

\section{Methodology}
All calculations are implemented in the Vienna \textit{ab initio} simulation package (VASP) based on the density functional theory (DFT) method \cite{vasp}. The Perdew, Burke, and Ernzerhof (PBE) parametrization within the generalized gradient approximation (GGA) is used for the exchange-correlation functional. A plane-wave basis set is employed with the kinetic energy cutoff of 600 eV. A 15$\times$15$\times$1 \textbf{k}-mesh is used during structural relaxation for the unit cell until the energy differences are converged within 10$^{-6}$ eV, with a Hellman-Feynman force convergence threshold of 10$^{-4}$ eV/\AA. We maintain the interlayer vacuum spacing larger than 12 \AA\ to eliminate the interaction with periodic boundary condition.

In the calculation of phonon dispersion, the harmonic interatomic force constants (IFCs) are obtained by density functional perturbation theory (DFPT) using the supercell approach, which calculates the dynamical matrix through the linear response of electron density \cite{DFPT}. The 5$\times$5$\times$1 supercell with 5$\times$5$\times$1 \textbf{k}-mesh is used to ensure the convergence. The phonon dispersion is obtained using the Phonopy code with the harmonic IFCs as input \cite{phonopy}.

\section{Results and discussion}
\subsection{Crystal structures}

\begin{figure}[h]
\centering
\includegraphics[width=\linewidth]{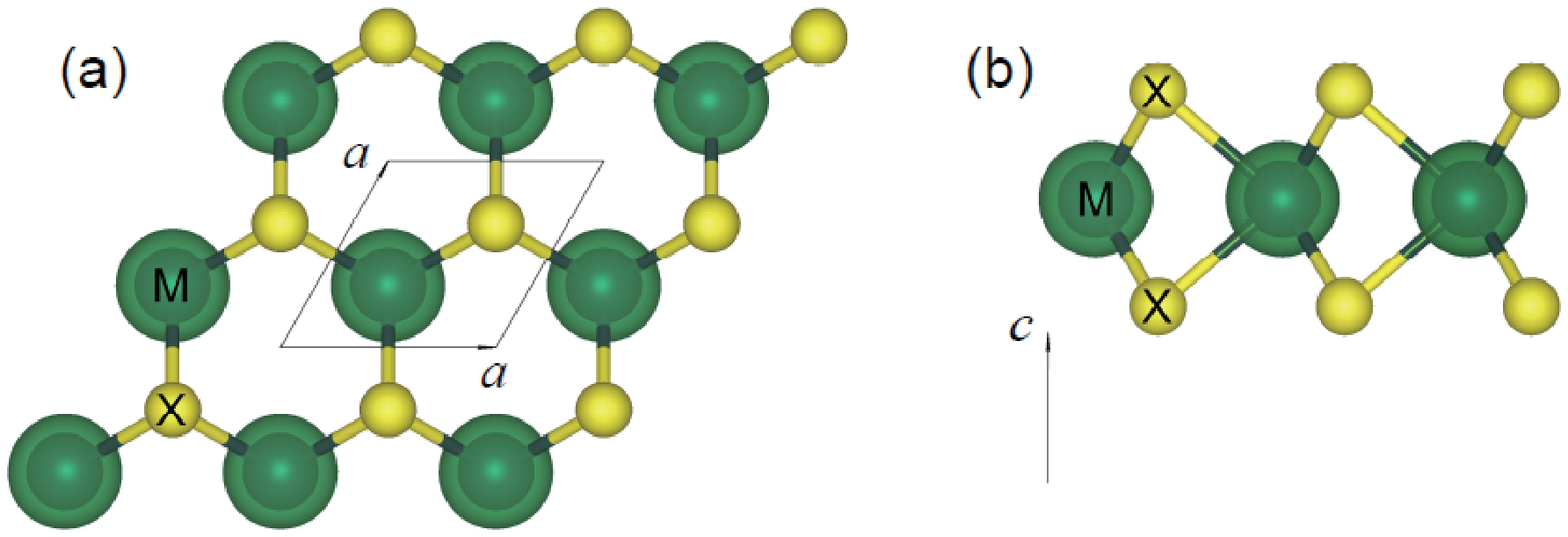}
\caption{(a) Top view and (b) side view of crystal structure of monolayer MX$_2$. The primitive vectors are \textbf{a}($a$,0,0), \textbf{b}($a/2$,$\sqrt{3}a/2$,0), and \textbf{c}(0,0,$c$).}
\label{structure} 
\end{figure}

Monolayer MX$_2$ has honeycomb structure with space group $P6m2$ \cite{MoS2-phonon} as shown in fig.~\ref{structure}. An M atom layer is sandwiched between two X atom layers, connected by covalent bonds. The optimized lattice parameters of all studied MX$_2$ are shown in table~\ref{lattice}. Our GGA calculations overestimate the lattice parameters by 0.16\%, 0.36\%, and 0.29\%, respectively, which is a general feature of the GGA functional.

\begin{table}
\centering
\caption{Calculated lattice parameters and band gap of monolayer MX$_2$.Experimental data are also given in parentheses for comparison.}
\begin{tabular}{llll}
\hline
 & MoS$_2$ & MoSe$_2$ & WS$_2$\\
\hline
$a$ (\AA) & 3.165  (3.160 $^\mathrm{a}$) & 3.300  (3.288 $^\mathrm{a}$) & 3.163  (3.154 $^\mathrm{b}$)\\
$E_g$ (eV) & 1.81  (1.88 $^\mathrm{c}$) & 1.56  (1.57 $^\mathrm{d}$) & 1.97 (1.95 $^\mathrm{e}$)\\
\hline
\multicolumn{4}{l}{$^\mathrm{a}$ Reference \cite{Coehoorn1987}}\\
\multicolumn{4}{l}{$^\mathrm{b}$ Reference \cite{WS2-a}}\\
\multicolumn{4}{l}{$^\mathrm{b}$ Reference \cite{Mak2010}}\\
\multicolumn{4}{l}{$^\mathrm{b}$ Reference \cite{Tonndorf:13}}\\
\multicolumn{4}{l}{$^\mathrm{b}$ Reference \cite{doi:10.1021/nl3026357}}\\
\end{tabular}
\label{lattice}
\end{table}

\begin{figure}[h]
\centering
\includegraphics[width=\linewidth]{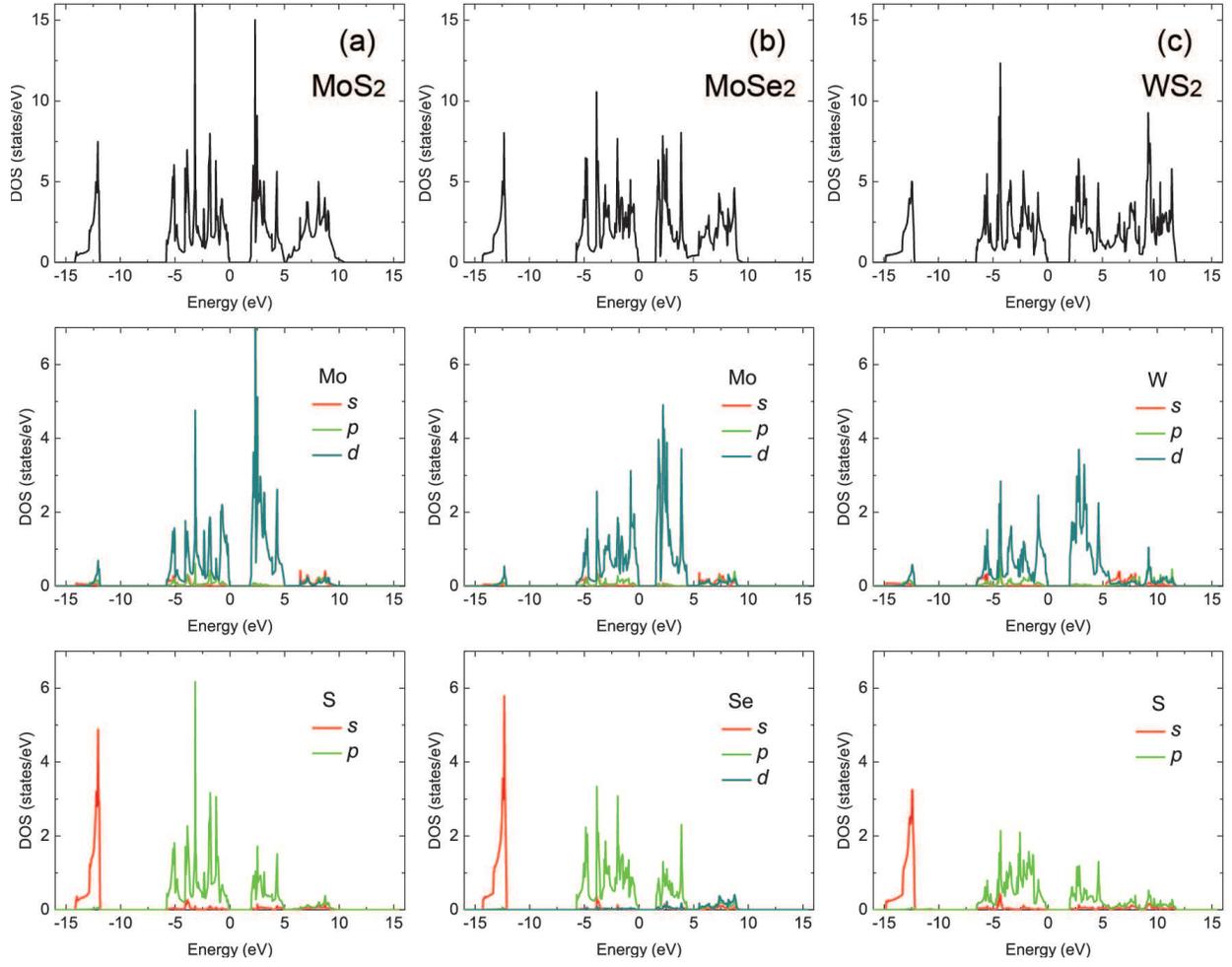}
\caption{Total and atom projected DOS for (a)$\mathrm{MoS}_2$, (b)$\mathrm{MoSe}_2$, and (c)$\mathrm{WS}_2$.}
\label{dos} 
\end{figure}

The electronic structures of all studied MX$_2$ are calculated by DFT method. As shown in table~\ref{lattice}, the calculated band gap is consistent with the measurements \cite{Mak2010,Tonndorf:13,doi:10.1021/nl3026357}. The total and atom projected density of states (DOS) are shown in fig.~\ref{dos}. The total DOS from -7 to 10 eV is mainly composed of M-$d$ and X-$p$ states, and the bands on each side of the band gap originate primarily from M-$d$ states, which is in agreement with previous work \cite{Kumar2012}. Due to a less localized DOS of W atoms, the overlap between W-$d$ and S-$p$ state in the valence band of WS$_2$ from -7 to 0 eV is larger than other two materials , indicating a strong covalent $p-d$ bonding. Similar large overlap between W-$d$ and S-$p$ state in WS$_2$ from 0 to 12 eV tends to increase the widths of the conduction bands.

\begin{figure}[h]
\centering
\includegraphics[width=0.6\linewidth]{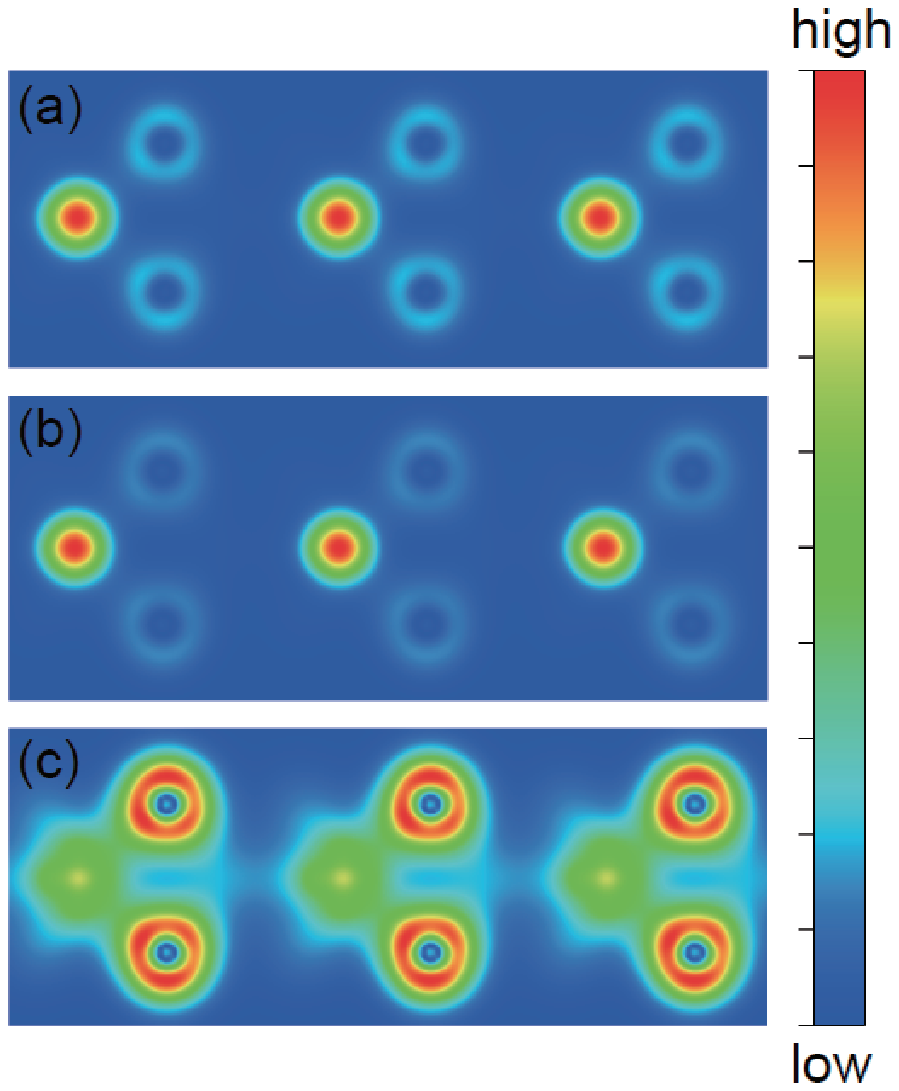}
\caption{Electronic charge density of (a)$\mathrm{MoS}_2$, (b)$\mathrm{MoSe}_2$, and (c)$\mathrm{WS}_2$ in the [$1\overline{1}0$] plane.}
\label{charge} 
\end{figure}

Fig.~\ref{charge} presents the electronic charge density of all studied MX$_2$ in the [$1\overline{1}0$] plane. For MoS$_2$ and MoSe$_2$, the highest charge density is found to be on the Mo atoms due to the strongly localized DOS of Mo atoms, while for WS$_2$, S atoms have the highest charge density. As shown in fig.~\ref{charge}, the W-S bonding is the strongest, whereas the Mo-Se bonding is weaker than the Mo-S bonding, which is consistent with the projected DOS in fig.~\ref{dos}. Usually, strong interatomic bonding and low average atomic mass imply a large Debye temperature, leading to a high thermal conductivity. From fig.~\ref{dos}, it is found that the $d$ state of transition metal M play an important role in determining the interatomic bonding, which will further affect the heat transport in these three materials.

\subsection{Phonon spectra}
\begin{figure}[h]
\centering
\includegraphics[width=0.9\linewidth]{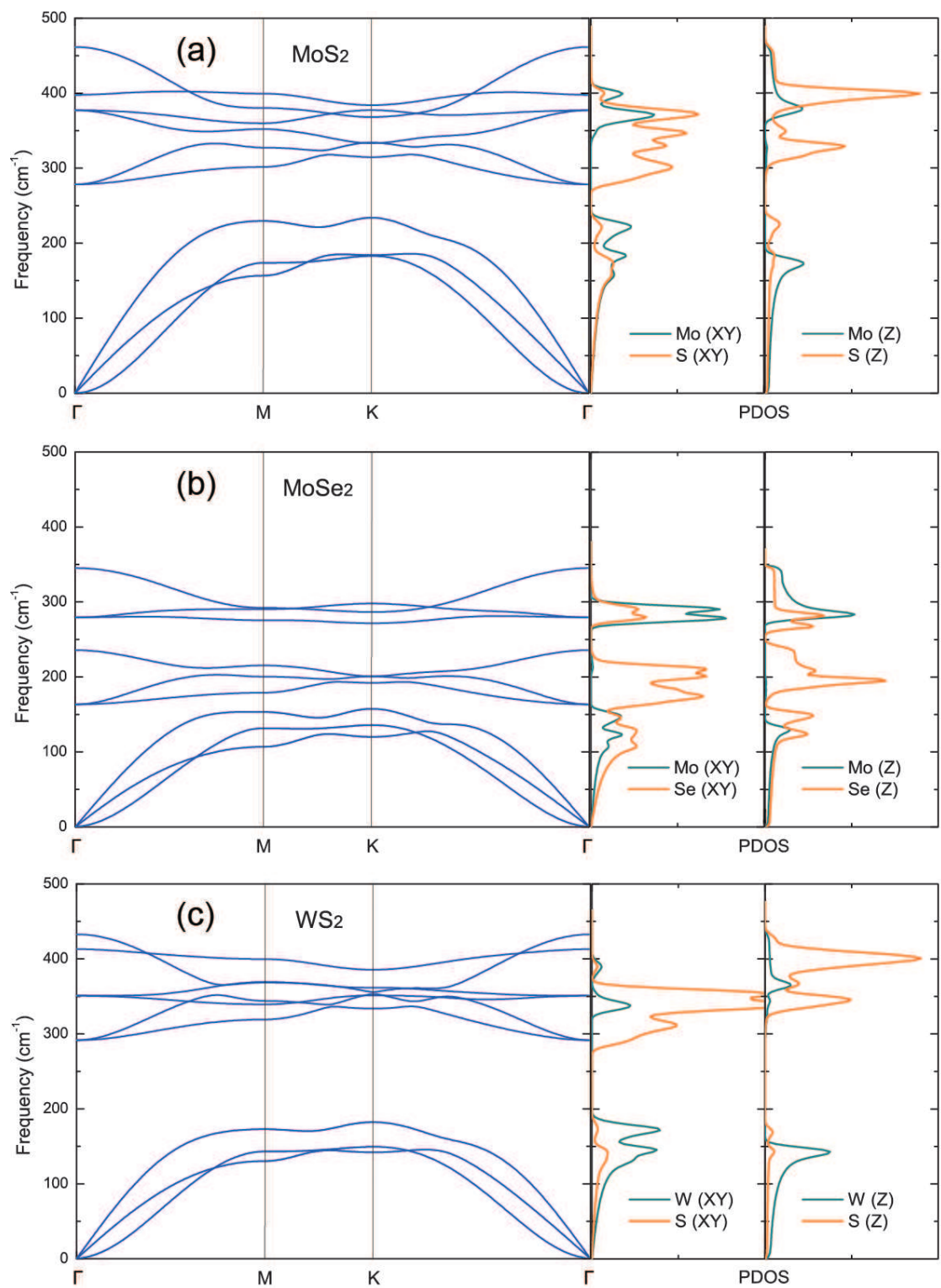}
\caption{Phonon spectra and projected PDOS for (a)MoS$_2$, (b)MoSe$_2$, and (c)WS$_2$.}
\label{phonon} 
\end{figure}

The phonon spectra of all studied MX$_2$ structures are calculated using the supercell approach, with the real-space force-constants calculated in the density-functional perturbation theory (DFPT) \cite{DFPT} within the Phonopy code \cite{phonopy}. Fig.~\ref{phonon} presents the phonon spectra along several high symmetry directions, together with the corresponding projected phonon density of states (PDOS). The primitive cell of monolayer MX$_2$ contains 3 atoms, corresponding to three acoustic and six optical phonon branches. The average acoustic Debye temperature for monolayer MX$_2$ is determined from \cite{Morelli2002}

\begin{equation}
\frac{1}{\theta_D^3}=\frac{1}{2}(\frac{1}{\theta_{\mathrm{LA}}^3}+\frac{1}{\theta_{\mathrm{TA}}^3}),
\end{equation} 

\noindent where $\theta_i$ for each acoustic branch $i$ ($i=$ LA, TA) is defined as 

\begin{equation}
\theta_i=\frac{\hbar \omega_{i,max}}{k_B},
\end{equation} 

\noindent where $\hbar$ is Planck constant, $\omega_{i,max}$ is the phonon frequency at the zone boundary of the $i$-th acoustic mode, and $k_B$ is Boltzmann constant. The calculated Debye temperature $\theta_D$ for MoS$_2$, MoSe$_2$, and WS$_2$ are 262.3 K, 177.6 K, and 213.6 K, respectively, which is in good agreement with previous results, $i.e.$ 260-320 K for MoS$_2$ estimated from specific-heat measurement \cite{debye-260-MoS2}, 197.3$\pm$6.6 K for MoSe$_2$ estimated from photoconductivity measurements \cite{Hu2006176}, 210 K for WS$_2$ estimated from the Lindemann formula \cite{debye-WS2}.

Concerning thermal vibrations and the bonding forces, the Debye temperature is a measure of the temperature above which all modes begin to be excited and below which modes begin to be frozen out \cite{debye-vibrations}. We first investigate the relation between vibrational modes and the Debye temperature by calculating the projected PDOSs for the M(XY), M(Z), X(XY), and X(Z) vibrations in MX$_2$ stuctures as shown in fig.~\ref{phonon}. Similar to the diatomic linear chain model, the scale of the acoustic (optical) phonon branch is dominated by atoms with larger (smaller) mass in three materials. As the mass ratio of all studied MX$_2$ ($m_{\mathrm{M}}/m_{\mathrm{2X}}$) in table~\ref{debyemass} show, the acoustic phonon vibration in the PDOS is governed by the larger mass. The mass ratio of MoS$_2$ is most close to 1, while that of WS$_2$ is much larger 1. Therefore the low-frequency acoustic phonon branches of MoS$_2$ up to 233.9 cm$^{-1}$ are mainly from the Mo(XY), Mo(Z) and S(XY) vibrations due to similar mass, whereas those of WS$_2$ up to 182.3 cm$^{-1}$ are mainly from the W(XY) and W(Z) vibrations due to the much larger mass of W atoms. In contrast to other two materials, the mass of transition metal atoms in MoSe$_2$ is smaller than the mass of chalcogenide atoms. Thus, although all Mo(XY), Mo(Z), Se(XY), and Se(Z) vibrations contribute significantly to the low-frequency branches of MoSe$_2$ up to 157.5 cm$^{-1}$, the PDOS of the Se(XY) and Se(Z) vibrations is higher than other two vibrations due to the relatively larger
mass of Se atoms.

Furthermore, low average atomic mass $\overline{M}$, besides the strong interatomic bonding mentioned above, can lead to a large Debye temperature as well. The mass differences in MoS$_2$, MoSe$_2$, and WS$_2$ are also shown in table~\ref{debyemass}. For MoS$_2$ and MoSe$_2$ which have similar bonding characteristics as shown in fig.~\ref{charge}, the $\overline{M}$ of MoS$_2$ is approximately two thirds the $\overline{M}$ of MoSe$_2$, and the calculated $\theta_{D}$ is about 1.5 times larger. For WS$_2$, although it has close mass to MoSe$_2$, the strong covalent W-S bonding as shown in fig.~\ref{charge}, can result in relatively larger Debye temperature as implied in previous work \cite{Lindsay2013}. Therefore, although the average atomic mass plays a key role in determining the Debye temperature, the effect of interatomic bonding can not be neglected and should be taken into account when studying the thermal transport properties of monolayer MX$_2$.

\begin{table}
\centering
\caption{The mass ratio $m_{\mathrm{M}}/m_{\mathrm{2X}}$, average atomic mass $\overline{M}$,and Debye temperature $\theta_{D}$ of all studied MX$_2$.}
\begin{tabular}{llll}
\hline
Structure & $m_{\mathrm{M}}/m_{\mathrm{2X}}$ & $\overline{M}$ (amu) & $\theta_{D}$ (K)\\
\hline
MoS$_2$ & 1.50 & 53.36 & 262.3\\
MoSe$_2$ & 0.61 & 84.63 & 177.6\\
WS$_2$ & 2.87 & 82.66 & 213.6\\
\hline
\end{tabular}
\label{debyemass}
\end{table}

In addition, we investigate the vibrational mode of all studied MX$_2$. Since the monolayer MX$_2$ belongs to the $D_{3h}$ point group, the optical lattice-vibration modes at $\Gamma$ point can be thus decomposed as

\begin{equation}
\Gamma_{optical}=A_2''(\mathrm{IR})+A_1'(\mathrm{R})+E'(\mathrm{IR+R})+E''(\mathrm{R}),
\label{opt-mode}
\end{equation}

\noindent where IR and R denote infrared- and Raman-active mode respectively. Table~\ref{phononvib} lists the optical phonon frequencies at the $\Gamma$ point. The calculated phonon frequencies are in agreement with the experimental results, and the discrepancy is less than 3\%. The LO/TO splitting is very small and can be neglected here \cite{Cai2014,Luttrell2006}.

\begin{table}
\centering
\caption{Theoretical determined optical phonon frequencies (cm$^{-1}$) at the $\Gamma$ point. The experimental results are also given in parentheses for comparison.}
\begin{tabular}{lllll}
\hline
Structure & $E''$ & $E'$ & $A_1'$ & $A_2''$\\
\hline
MoS$_2$ & 278.4 (283 $^\mathrm{a}$) & 377.2 (385 $^\mathrm{a}$) & 397.7 (404 $^\mathrm{a}$) & 461.6 (470 $^\mathrm{b}$)\\
MoSe$_2$ & 163.3 (167 $^\mathrm{c}$) & 235.7 (240 $^\mathrm{c}$) & 279.3 (282 $^\mathrm{c}$) & 345.2 (351 $^\mathrm{c}$)\\
WS$_2$ & 291.6 (298 $^\mathrm{a}$) & 350.8 (357 $^\mathrm{a}$) & 413.0 (419 $^\mathrm{a}$) & 432.8 (438 $^\mathrm{d}$)\\
\hline
\multicolumn{5}{l}{$^\mathrm{a}$ Reference \cite{C4CS00282B}}\\
\multicolumn{5}{l}{$^\mathrm{b}$ Reference \cite{Wieting1971}}\\\multicolumn{5}{l}{$^\mathrm{c}$ Reference \cite{Late2014}}\\
\multicolumn{5}{l}{$^\mathrm{d}$ Reference \cite{Luttrell2006}}\\
\end{tabular}
\label{phononvib}
\end{table}

\begin{figure}[h]
\centering
\includegraphics[width=\linewidth]{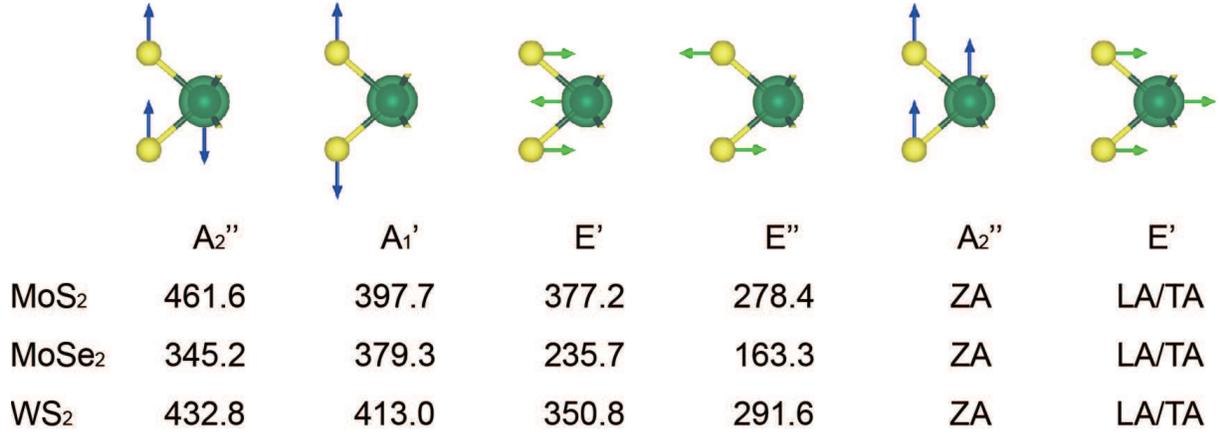}
\caption{Schematic phonon vibrations with different frequencies (cm$^{-1}$).}
\label{mode} 
\end{figure}

The schematic vibrations for the phonon modes are shown in fig.~\ref{mode}, where one $A_2''$ and one $E'$ are acoustic modes, the other $A_2''$ ($E'$) are IR (both IR and R) active as shown in Eq.~(\ref{opt-mode}). $A_2''$ and $A_1'$ modes vibrate along the $z$-direction, and $E'$ and $E''$ modes vibrate in the $x-y$ direction. As shown before in fig.~\ref{phonon}, in the case of $E'(\mathrm{LA/TA})$ in monolayer MoS$_2$, the Mo and S atoms vibrate with similar amplitudes; for $A_2''(\mathrm{ZA})$ in MoS$_2$, the vibrations of Mo atoms have much larger amplitudes. For both $E'(\mathrm{LA/TA})$ and $A_2''(\mathrm{ZA})$ in monolayer MoSe$_2$, the Se atoms vibrate with greater amplitudes than Mo atoms. The vibration of W atoms dominates S atoms in both $E'(\mathrm{LA/TA})$ and $A_2''(\mathrm{ZA})$ vibrational modes in monolayer WS$_2$ due to the large $m_{\mathrm{W}}/m_{\mathrm{2S}}$.

\subsection{Thermodynamic properties}

The Gr\"uneisen parameter $\gamma$, which describes the thermal expansion of a crystal on its vibrational properties, provides information on the anharmonic interactions. The larger the Gr\"uneisen parameter indicates the stronger anharmonic vibrations. The expression for the Gr\"uneisen parameter is given by \cite{Grimvall,shao2015}

\begin{equation}
\gamma= \frac{3\alpha BV_m}{C_V},
\label{parameter}
\end{equation}
where $\alpha$ is the linear thermal expansion coefficient, $B$ is the bulk modulus, $V_m$ is the molar volume, and $C_V$ is the isometric heat capacity.

Table~\ref{thermal} compares the calculated isometric heat capacity, bulk modulus, and linear thermal expansion coefficient with experimental results at 300 K. The isometric heat capacity can be calculated as

\begin{equation}
{C_V}=\left(\frac{\partial E}{\partial T}\right)_V= \sum_{n,\textbf{q}}k_B\left(\frac{\hbar\omega_n(\textbf{q})}{k_BT}\right)^2\frac{e^{\hbar\omega_n(\textbf{q})/k_BT}}{(e^{\hbar\omega_n(\textbf{q})/k_BT}-1)^2},
\end{equation} 

\noindent where $T$ is temperature, and $\omega_n(\textbf{q})$ is the phonon frequency of the $n$-th branch with wave vector $\textbf{q}$. The calculated values of $C_V$ for $\mathrm{MoS}_2$, $\mathrm{MoSe}_2$, and $\mathrm{WS}_2$ at room temperature are in good agreement with the experimental results \cite{MoS2-Cv,MoSe2-Cv,WS2-Cv}. The bulk modulus $B$ and linear thermal expansion coefficient $\alpha$ are calculated using the quasi-harmonic approximation (QHA), which takes the first-order anharmonicity into account \cite{phonopy}. The obtained $B$ for $\mathrm{MoS}_2$ is in agreement with the experimental value \cite{MoS2-B}. For $\mathrm{MoSe}_2$, the computed $B$ is fallen in the range of measured values \cite{MoSe2-B,MoSe2-B2}. The great discrepancies between the calculated and experimental $B$ are seen for $\mathrm{WS}_2$. Since the bulk modulus is used to describe the stiffness of MX$_2$ \cite{Huang2015}, the bonding in monolayer WS$_2$ is found to be much stiffer comparing to other two materials. The calculated $\alpha$ for MoS$_2$ and MoSe$_2$ at 300 K is is fallen in the range of measured values \cite{Late2014,MoS2-a,MoSe2-a}, while larger than the measured $\alpha$ for WS$_2$.

\begin{table}
\centering
\caption{Comparison between the calculated and measured isometric heat capacity $C_V$ (J mol$^{-1}$ K$^{-1}$), bulk modulus $B$ (GPa), and linear thermal expansion coefficient $\alpha$ (10$^{-6}$ K$^{-1}$). The experimental results are also given in parentheses.}
\begin{tabular}{llll}
\hline
Structure & $C_V$ & $B$ & $\alpha$\\
\hline
MoS$_2$ & 62.97 (63.55 $^\mathrm{a}$) & 52.3 (53.4$\pm$1.0 $^\mathrm{b}$) & 17.4 (10.7 $^\mathrm{c}$) (82 $^\mathrm{d}$)\\
MoSe$_2$ & 68.75 (68.60 $^\mathrm{e}$) & 57.3 (45.7$\pm$0.3 $^\mathrm{f}$) (62 $^\mathrm{g}$) & 19.5 (7.24 $^\mathrm{h}$) (105 $^\mathrm{d}$)\\
WS$_2$ & 63.49 (63.82$\pm$0.32 $^\mathrm{i}$) & 77.9 (61$\pm$1 $^\mathrm{j}$) & 14.8 (6.35 $^\mathrm{k}$)\\
\hline
\multicolumn{4}{l}{$^\mathrm{a}$ Reference \cite{MoS2-Cv}}\\
\multicolumn{4}{l}{$^\mathrm{b}$ Reference \cite{MoS2-B}}\\
\multicolumn{4}{l}{$^\mathrm{c}$ Reference \cite{MoS2-a}}\\
\multicolumn{4}{l}{$^\mathrm{d}$ Reference \cite{Late2014}}\\
\multicolumn{4}{l}{$^\mathrm{e}$ Reference \cite{MoSe2-Cv}}\\
\multicolumn{4}{l}{$^\mathrm{f}$ Reference \cite{MoSe2-B}}\\
\multicolumn{4}{l}{$^\mathrm{g}$ Reference \cite{MoSe2-B2}}\\
\multicolumn{4}{l}{$^\mathrm{h}$ Reference \cite{MoSe2-a}}\\
\multicolumn{4}{l}{$^\mathrm{i}$ Reference \cite{WS2-Cv}}\\
\multicolumn{4}{l}{$^\mathrm{j}$ Reference \cite{WS2-B}}\\
\multicolumn{4}{l}{$^\mathrm{k}$ Reference \cite{WS2-a}}\\
\end{tabular}
\label{thermal}
\end{table}

\begin{figure}[h]
\centering
\includegraphics[width=\linewidth]{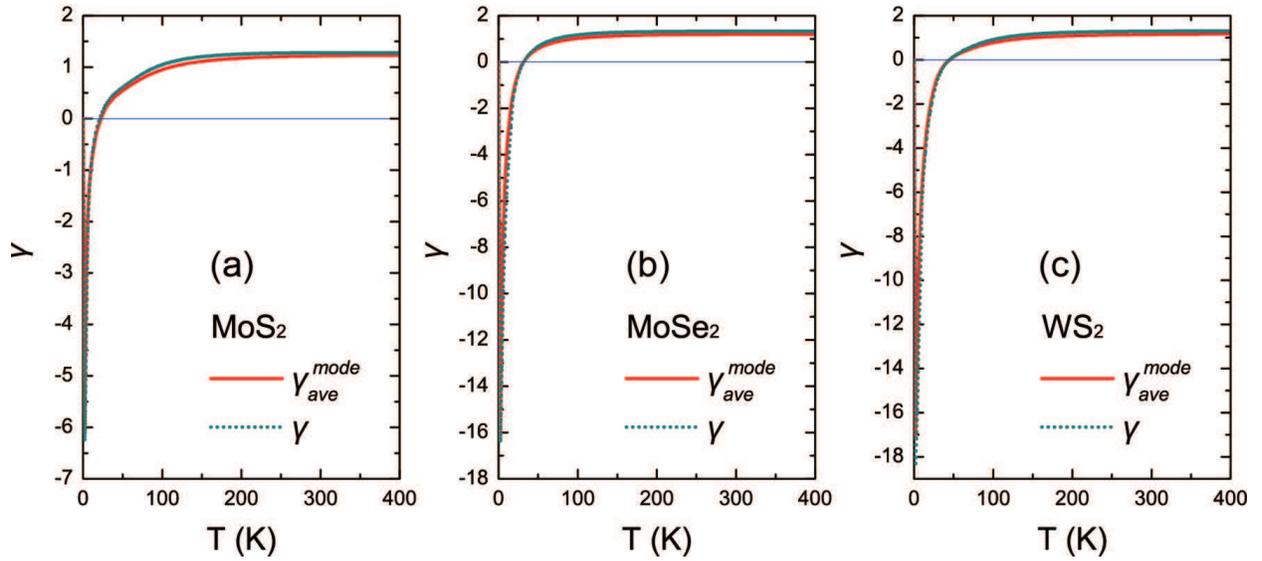}
\caption{Calculated temperature-dependent Gr\"uneisen parameter for (a)$\mathrm{MoS}_2$, (b)$\mathrm{MoSe}_2$, and (c)$\mathrm{WS}_2$.}
\label{temperature} 
\end{figure}

As shown in fig.~\ref{temperature}, the temperature-dependent Gr\"uneisen parameter is calculated using Eq.~(\ref{parameter}). The Gr\"uneisen parameter can also be calculated by averaging the mode Gr\"uneisen parameter $\gamma_n(\textbf{q})$,

\begin{equation}
\gamma_{ave}^{mode}= \frac{1}{C_V}\sum_{n,\textbf{q}}\gamma_n(\textbf{q})C_{V,n}(\textbf{q}),
\end{equation}

\noindent where $C_{V,n}(\textbf{q})$ is the mode heat capacity. The mode Gr\"uneisen parameter is given by

\begin{equation}
\gamma_n(\textbf{q})=-\frac{a_0}{\omega_n(\textbf{q})}\frac{\partial{\omega_n(\textbf{q})}}{\partial{a}},
\end{equation}

\noindent where $a_0$ is the equilibrium lattice constant at 0 K. Fig.~\ref{temperature} also shows the calculated $\gamma_{ave}^{mode}$, which is consistent with the Gr\"uneisen parameter calculated using Eq.~(\ref{parameter}).

\begin{figure}[h]
\centering
\includegraphics[width=0.6\linewidth]{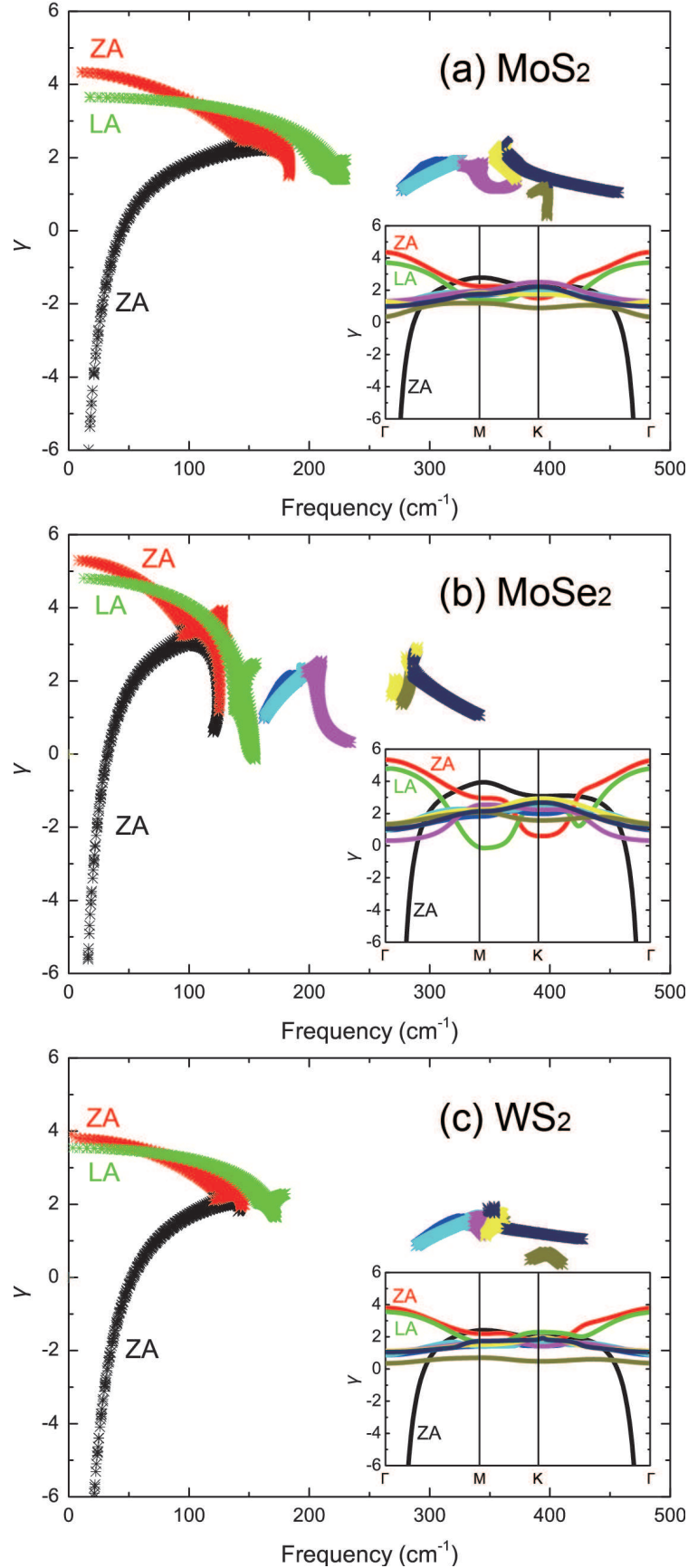}
\caption{Calculated mode Gr\"uneisen parameter for (a)MoS$_2$, (b)MoSe$_2$ and (c)WS$_2$ with respect to frequencies, and with respect to wave vectors as shown in the inset.}
\label{mode-gruneisen} 
\end{figure}

The frequency dependence of mode Gr\"uneisen parameter of MoS$_2$ in the irreducible BZ is plotted in fig.~\ref{mode-gruneisen}, and the mode Gr\"uneisen parameter along symmetry directions is shown in the inset. Similar to diamond, graphite and graphene \cite{Mounet2005}, negative $\gamma_{\mathrm{ZA}}$ is observed at low frequencies (under 50 cm$^{-1}$ in MoS$_2$). The average Gr\"uneisen parameter is negative at low temperatures because only the low-frequency phonons are excited, as shown in fig.~\ref{temperature}. Our results are in agreement with previous calculations \cite{Huang2014,Sevik2014}. The calculated $\gamma_{ave}^{mode}$ at room temperature is 1.22, 1.20, and 1.15 for MoS$_2$, MoSe$_2$, and WS$_2$, respectively, indicating that MoS$_2$ has the strongest bonding anharmonicity among three materials, while WS$_2$ has the weakest anharmonic vibrations. The bonding anharmonicity obtained from analysing the $\gamma$ of all studied MX$_2$ is consistent with previous work \cite{Huang2015}.

According to Slack's expression \cite{Slack1973321,Morelli2002}, assuming that only the acoustic phonon modes participate in the heat conduction process, the lattice thermal conductivity in the temperature range where three-phonon scattering is dominant, can be given as

\begin{equation}
\kappa_l= A\frac{\overline{M}\theta_D^3\delta n^{1/3}}{\gamma^2T},
\label{slack}
\end{equation}

\noindent where $\overline{M}$ is the average mass of an atom in the crystal, $\delta^3$ is the volume per atom, $n$ is the number of atoms in the primitive unit cell, and $A$ is a constant which is given by \cite{Julian1965}

\begin{equation}
A= \frac{2.43\times10^{-6}}{1-0.514/\gamma+0.228/\gamma^2}
\end{equation}

\noindent when the units of $\kappa_l$, $\overline{M}$, and $\delta$ are taken as W/mK, amu, and \AA, respectively. The obtained lattice thermal conductivity for monolayer MoS$_2$, MoSe$_2$, and WS$_2$ at room temperature is 33.6 W/mK, 17.6 W/mK, and 31.8 W/mK, respectively, which is in good agreement with the experimental value of $34.5\pm4$ W/mK for monolayer MoS$_2$ \cite{monoMoS2-k}, and 32 W/mK for monolayer WS$_2$ \cite{raey}. Although there is no experimental value for monolayer MoSe$_2$, our calculated $\kappa_l$ is a reasonable prediction.  

The Slack's expression attempts to normalize the effect of mass density, crystal structure, interatomic bonding, and anharmonicity \cite{Slack1973321,Lindsay2013,Jain2014}. MoS$_2$, MoSe$_2$, and WS$_2$ have similar crystal structure. The factor $\overline{M}\theta_D^3\delta$ in Eq.~(\ref{slack}) is maximized for light mass, strong bonded crystals, because low average atomic mass and strong interatomic bonding lead to a high $\theta_D^3$, and the $\theta_D^3$ term dominates the behaviour \cite{Slack1973321}. The Debye temperature reflects the magnitude of sond velocity. Higher Debye temperature results in increased phonon velocities, and increased acoustic-phonon frequencies as mentioned above, which suppress phonon-phonon scattering by decreasing phonon populations. Therefore the high thermal conductivity of MoS$_2$ is related to its high Debye temperature, which is due to its much lower average atomic mass.

Recent theoretical work has predicted that monolayer WS$_2$ has the highest thermal conductivity among all studied MX$_2$ at room temperature due to a large frequency gap between its acoustic and optic phonons \cite{Gu2014}, which originates in its larger mass ratio $m_{\mathrm{W}}/m_{\mathrm{2S}}$. Our results suggest that the average atomic mass plays a key role in determining the phonon dispersion of MoS$_2$ and WS$_2$, and subsequently determines the Debye temperature from which the phonon velocities can be estimated. In addition, although WS$_2$ has similar average atomic mass to MoSe$_2$, its strong W-S bonding leads to a higher Debye temperature. Furthermore, small $\gamma$ of WS$_2$ means low anharmonicity, which also results in a higher thermal conductivity. Therefore the thermal conductivity of WS$_2$ is determined by the competition between high average atomic mass, strong covalent W-S bonding and low anharmonicity.

\section{Conclusion}
In summary, we investigate the lattice dynamics and thermodynamic properties of MoS$_2$, MoSe$_2$, and WS$_2$ by first principles calculations. The obtained phonon frequencies and lattice thermal conductivity agree well with experimental measurements. Our calculations show that the thermal conductivity of MoS$_2$ is highest among the three materials due to its large Debye temperature, which is attributed to the lowest average atomic mass. We also find that WS$_2$ has stronger covalent W-S bonding and lower anharmonicity, leading to much higher thermal conductivity compared to MoSe$_2$.

\section*{Acknowledgement}
This work is supported by the National Natural Science Foundation of China under Grants No. 11374063 and 11404348.

\clearpage

\end{document}